\documentclass[aps,prl,twocolumn,groupedaddress,nofootinbib]{revtex4-1}
\usepackage{amsmath}
\usepackage{graphicx}
\usepackage{sistyle}
\usepackage{bm}
\usepackage{natbib}
\usepackage{dsfont}
\usepackage{epsfig}
\usepackage{feynmf}
\usepackage{blindtext, rotating}
\usepackage{mathtools}
\usepackage{dsfont}
\usepackage{subcaption}
\usepackage{bbold}
\usepackage{amsfonts}
\usepackage{ragged2e}

\DeclareCaptionJustification{justified}{\justifying}

\begin{document}
\title{Quantum Molecular Robots}

\author{Thiago Guerreiro}
\email{barbosa@puc-rio.br}

\affiliation{Department of Physics, Pontifical Catholic University of Rio de Janeiro, Rio de Janeiro 22451-900, Brazil}

\begin{abstract}
Living organisms exploit complex molecular machines to execute crucial functions in chaotic environments. Inspired by nature's molecular setups we explore the idea of a quantum mechanical device whose purpose is self-protecting quantum information against a noisy environment.
\end{abstract}

\maketitle

\section{Introduction}

Molecules that execute tasks in crowded noisy environments are pervasive in biology. 
Molecular machines, as they are often called, are complex structures with thousands of atoms self-orchestrated to perform specific functions \cite{Goodsell1993} that draw energy and information from their surroundings to sustain motional and chemical non-equilibrium states essential to life \cite{Schrodinger, Astumian1998}. 
It is often discussed whether quantum mechanical effects play a role in these phenomena, and it has many times been suggested that coherence and entanglement can improve the efficiency of bio-molecular processes \cite{
Engel2007, Plenio2008, Svetlichny2012, Shabani2012, Walters2014, Svetlichny2017}. 
We propose an exploration of the \textit{dual} question: what lessons can we draw from biology that would improve control over quantum mechanical matter?

Nanotechnology has been largely inspired by nature's machines \cite{Feynman, Zhang2018}. Artificial molecular pumps \cite{Stoddart2015}, motors \cite{Kassem2017}, classical \cite{Lau2017}  and quantum tunneling ratchets \cite{Linke1999, Salger2009} and microscopic versions of Maxwell's demon \cite{Serreli2007} have been synthesized and studied \cite{Cheng2015}. Molecular data storage \cite{Ceze2019} and processing \cite{Bonnet2013} using DNA are also possible, and gene editing technology is now standard \cite{CRISPR}.
In parallel to developments in nanotechnology, the study of ultracold quantum chemistry \cite{Ospelkaus2010} has enabled the observation of molecular assembly from single atoms \cite{Liu2017} as well as chemical reactions at the single molecule level \cite{Hu2019}. 
Quantum interference of large organic molecules has been demonstrated \cite{Gerlich2011, Fein2019} and it has been argued that the technology for preparing large biological systems in quantum superposition states is within reach \cite{Isart2010}.

It is expected that eventually, ultracold quantum chemistry and nanotechnology will merge, perhaps fueled by the upcoming advances in quantum simulation and computation \cite{Arute2019}. 
One could then imagine the experimental realization of \textit{quantum molecular machines}, which we will refer to as quantum robots, or \textit{qubots} for short. 
Inspired by biological molecular machines such as kinesin \cite{Andreasson2015}, RNA Polymerase \cite{Chakraborty2016} and self-replicating ribosymes \cite{Paul2002}, qubots draw energy and information from their surroundings with the purpose of maintaining and processing quantum mechanical non-equilibrium states against the detrimental actions of decoherence and thermalization. Similar to naturally occurring machines, such structures have the potential of being significantly complicated, yet nothing prevents them from having molecular dimensions. A qubot can thus be thought of as an active self-error correcting molecule functioning as a quantum memory or processor.  

Engineering the environment and exploiting the dynamics of open systems to protect quantum states and achieve fault tolerance is an interesting idea with proposals in quantum information and optics \cite{Diehl2008, Verstraete2009, Vacanti2009, Reiter2012, Reiter2017}. Moreover, it is known that noise and decoherence are not always detrimental for entangled states \cite{Plenio1999, Plenio2001}, and that autonomous quantum thermal engines can prepare entanglement in the steady state \cite{Brask2015, Schulman1999}. 
Topological error correcting codes are also an example of naturally fault tolerant systems \cite{Kitaev2003, Bombin2013}.
Qubots differ from the idea of an engineered environment and topological codes as they actively draw energy and resources from the external world to protect their quantum data against a \textit{pre-existing} non-engineered environment. 
An interesting parallel can be drawn with a quantum error correcting code \cite{Fowler2012}, where syndrome measurements are made and correction operators are subsequently applied by an external agent. 
For a qubot, the error detection and correction are intrinsic parts of itself, occurring as a consequence of their natural non-equilibrium dynamics, in a similar manner DNA Polymerase performs proofreading and error correction on a strand of DNA \cite{Milo}. 

In this work we explore possibilities around the concept of quantum molecular machines. 
The main idea is to provide a simple toy example that captures the essence of a qubot.
In the following section we construct such an example. Next we give an effective master equation description of the model. A discussion on possible implementations and future perspectives follows.



\section{Example of a Qubot}

We give an example of a quantum robot that stabilizes one e-bit of quantum information in the form of a singlet state against a dephasing environment. 
The construction can be seen as an algorithm running on an \textit{open} quantum computer and we will assume the ability to execute two-qubit operations among subsystems. Natural units ($ \hbar~=~1 $) are used throughout.


\begin{figure*}[t] 
    \centering
    \includegraphics[width = 0.8\textwidth]{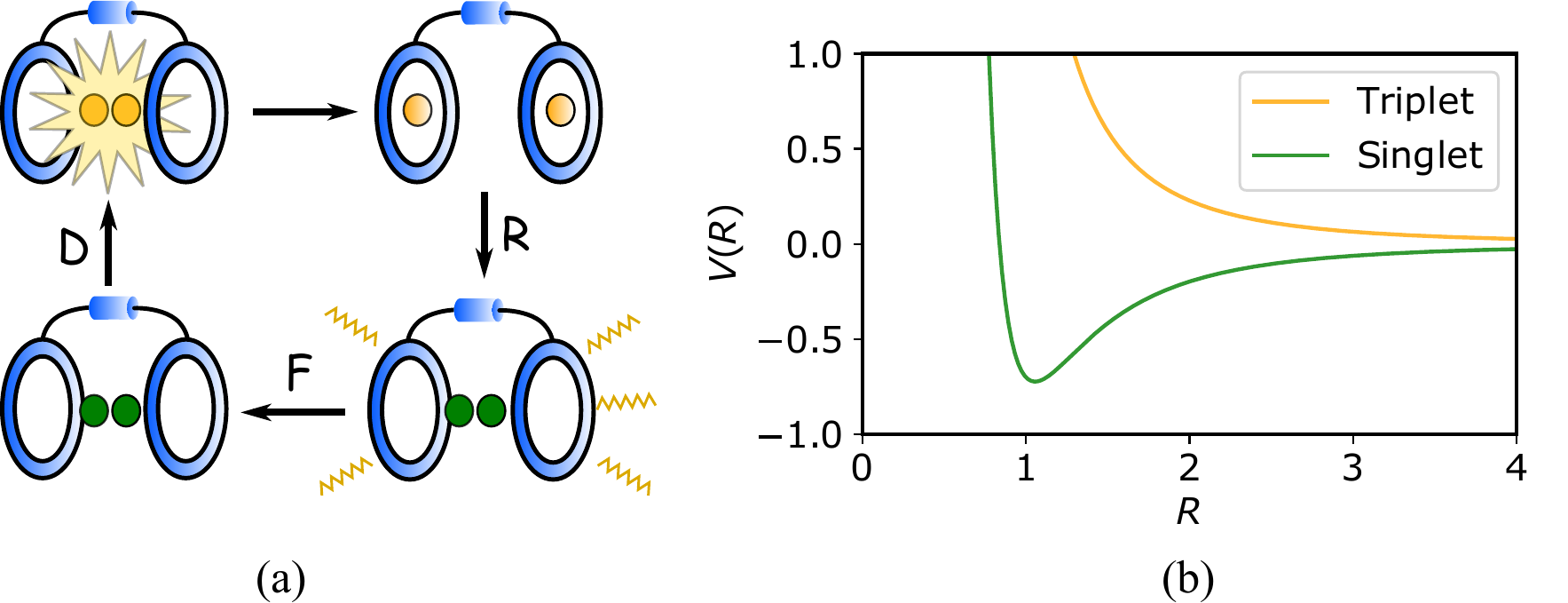}
    \caption[]{\small  (a) A ludic representation of the qubot and its cycle. Particles $ A, B $ are represented as spheres. (b) The potential between the protected spins.}\label{sketches}
    \label{color}
\end{figure*}

A ludic drawing of the qubot and its different configurations is shown in Figure \ref{sketches}(a). Consider a pair of particles $ A, B $. These particles will constitute the ``nucleous'' of the qubot and provide the \textit{protected} spin degrees of freedom. The particles have spin 1/2, and interact with each other via an ``interatomic'' potential of the form
\begin{eqnarray}
V(R) = V_{0}(R)+ \vec{S}_{A} \cdot \vec{S}_{B} \ V_{1}(R)
\end{eqnarray}
where $ R $ is the inter-particle distance and $ \vec{S}_{A}, \vec{S}_{B} $ the particles' spin operators.
The exact form of $ V_{0}(R) $ and $ V_{1}(R) $ is not important as long as the potential $ V(R) $ assumes a shape like the sketch shown in Figure \ref{sketches}(b) for the singlet $ \vert s \rangle $ and triplet $ \vert t \rangle $ states, defined as 
\begin{eqnarray}
\vert s \rangle &=& \dfrac{\vert \uparrow \rangle_{A} \vert \downarrow \rangle_{B}  - \vert \downarrow \rangle_{A} \vert \uparrow \rangle_{B} }{\sqrt{2}}  \\ \nonumber
\\
\vert t \rangle &=& \dfrac{\vert \uparrow \rangle_{A} \vert \downarrow \rangle_{B}  + \vert \downarrow \rangle_{A} \vert \uparrow \rangle_{B} }{\sqrt{2}}
\end{eqnarray}
Inspired by atomic physics we may say that $ V_{0}(R) \propto R^{-3} $ and $ V_{1}(R) \propto R^{-6} $, and particles $ AB$ form a molecular term $ ^{1}\Sigma^{+} $ for the singlet and $ ^{3}\Sigma^{+} $ for the triplet states \cite{Budker}.
%


Around the particles $ AB $ there is a ``superconducting circuit'' with a pair of loop shaped sensors which we will refer to as \textit{the loop} or simply as the letter $ L $.
Interaction between the particles and the circuit only becomes appreciable if they get very near, or inside the loops.
The circuit has two possible states denoted $ \vert \Phi_{0} \rangle $ and $ \vert \Phi_{1} \rangle $ with an energy gap $ \Delta $ between them and a free evolution determined by the Pauli operator $ H_{L} = \frac{\Delta }{2} Z $, where $ Z = \vert \Phi_{0} \rangle \langle \Phi_{0} \vert -  \vert \Phi_{1} \rangle \langle \Phi_{1} \vert   $. 
Through the loops, the circuit is sensitive to the particles' spins and is wired in such a way that when particle $ A $ enters the left loop and particle $ B $ enters the right loop the following operation is executed,
\begin{eqnarray}
\vert s \rangle \vert \Phi_{0} \rangle &\rightarrow & \vert s \rangle \vert \Phi_{0} \rangle \label{LAB1} \\
\vert t \rangle \vert \Phi_{0} \rangle &\rightarrow & \vert s \rangle \vert \Phi_{1} \rangle \label{LAB2}
\end{eqnarray}
Such operation could be for example a SWAP gate on the $ \lbrace \vert s \rangle \vert \Phi_{0} \rangle, \vert s \rangle \vert \Phi_{1} \rangle, \vert t \rangle \vert \Phi_{0} \rangle, \vert t \rangle \vert \Phi_{1} \rangle \rbrace $ basis. Details on how such coupling between superconducting qubits and spins can be achieved are not important for our discussion, but can be found in \cite{Marcos2010}. 
Note that this is an operation \textit{conditional} on the particles' position degree of freedom, and hence it induces a non-unitary operation, or dissipator, on the particles' spin and loop subspace. 

Now, suppose the environment decoheres the spin state of particles $ AB $  according to,
\begin{align}
\vert s \rangle_{AB} \vert 0 \rangle_{E} \vert \Phi_{0} \rangle_{L} &\,\mapsto \sqrt{1 - p} \vert s \rangle_{AB} \vert 0 \rangle_{E} \vert \Phi_{0} \rangle_{L} + \nonumber \\ \nonumber
\\ \nonumber
& + \sqrt{p} \left(  \dfrac{\vert \uparrow \rangle_{A} \vert \downarrow \rangle_{B} \vert 1 \rangle_{E} -  \vert \downarrow \rangle_{A} \vert \uparrow \rangle_{B} \vert 2 \rangle_{E}   }{\sqrt{2}}  \right) \vert \Phi_{0}  \rangle_{L}\\ \nonumber
\\ \nonumber
& =\sqrt{1 - p} \vert s \rangle_{AB} \vert 0 \rangle_{E} \vert \Phi_{0} \rangle_{L} + \\ \nonumber
\\ 
& + \sqrt{p} \left(  \dfrac{\vert s \rangle_{AB}  \vert + \rangle_{E} +  \vert t \rangle_{AB} \vert - \rangle_{E}   }{\sqrt{2}}  \right) \vert \Phi_{0} \rangle_{L} \label{decohered}
\end{align}
where $ \vert 0 \rangle_{E} , \vert 1 \rangle_{E}, \vert 2 \rangle_{E}$ are orthogonal states of the environment, $ p $ is the probability of error and $ \vert \pm \rangle_{E} = \left( \vert 1 \rangle_{E} \pm \vert 2 \rangle_{E} \right) / \sqrt{2} $. This channel can be seen as \textit{dephasing} of a logical qubit spanned by the states defined as $ \vert \uparrow \downarrow \rangle_{AB} \equiv \vert \bar{0} \rangle, \vert  \downarrow\uparrow \rangle_{AB} \equiv \vert \bar{1} \rangle  $.
We denote the dephasing superoperator acting on $ ABL $ as $ \mathcal{D} $. 
Dephasing is one example of an environment. For a different example, dimer destruction by photodissociation, see the Appendix.
After application of  $ \mathcal{D} $ there is a non-vanishing probability that the particles $ AB$ are found in a triplet state. 
In that case the interaction energy $V(R)$ between $ AB $ will change to a repulsive potencial and cause the particles to push each other apart and into the loops.

\begin{figure*}[t] 
    \centering
    \begin{subfigure}[b]{0.4\textwidth}
    \centering
        \includegraphics[width=\textwidth]{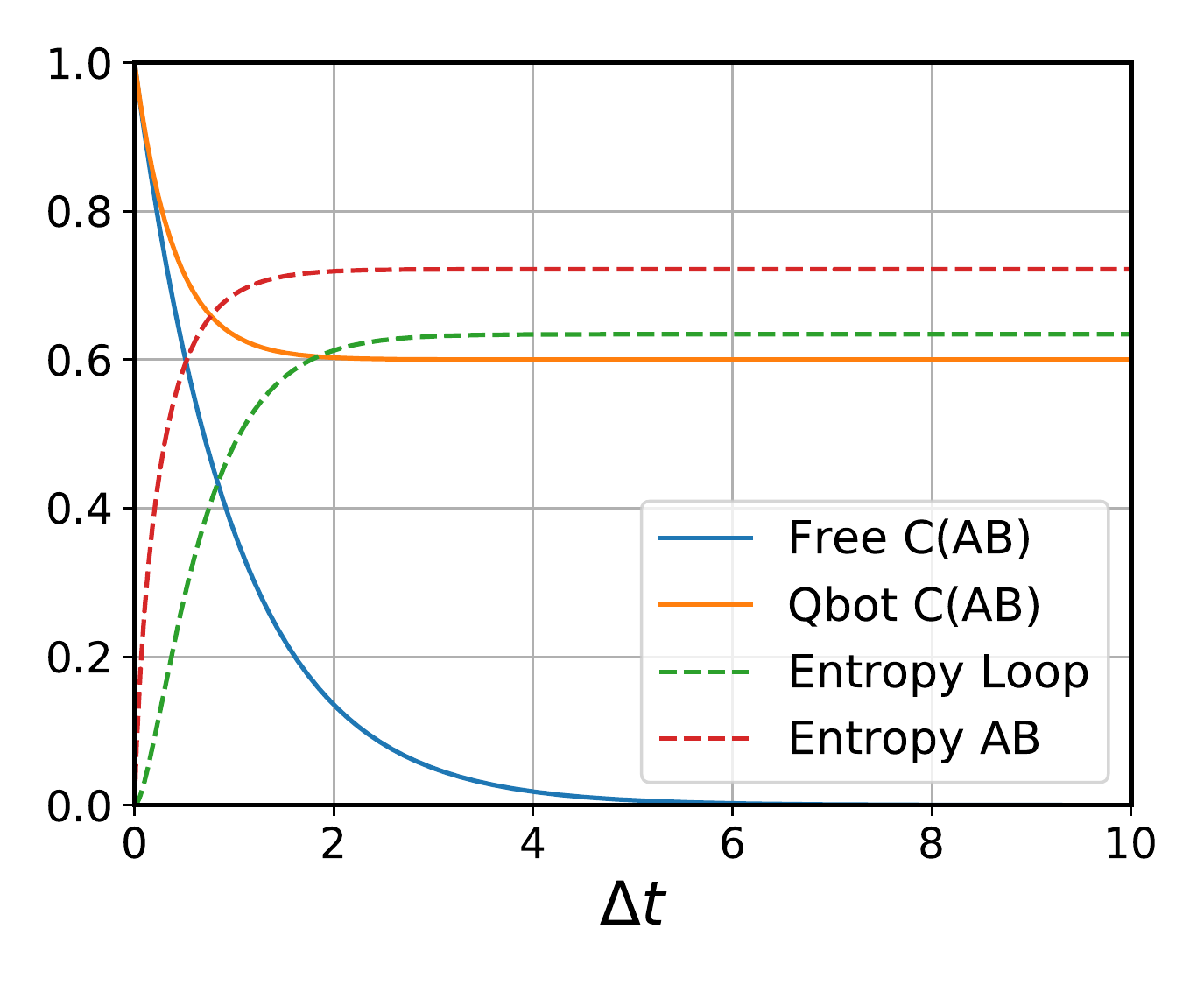}
        \caption{ \ }
        \label{fig:gull}
    \end{subfigure}
    ~ 
    \begin{subfigure}[b]{0.4\textwidth}
    \centering
        \includegraphics[width=\textwidth]{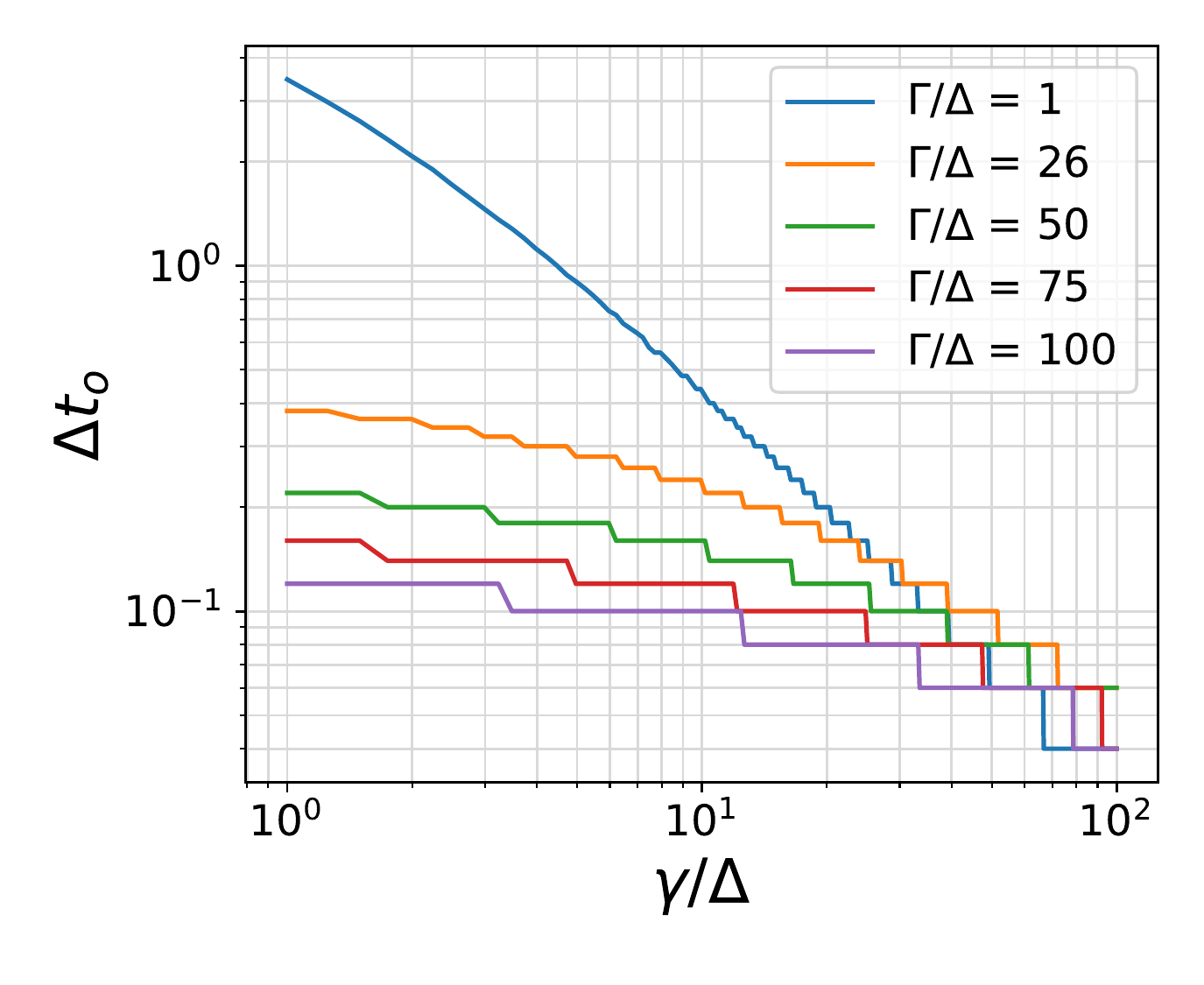}
        \caption{ \ }
        \label{fig:tiger}
    \end{subfigure}
    \caption[]{\small  (a) Transients of concurrence and entropy of $ AB$ and entropy of loop as a function of time measured in units of the inverse loop energy $ \hbar / \Delta $; rates for this plot are $ \Gamma / \Delta = 1, \gamma/\Delta = r / \Delta = 1.5 $. (b) Stabilization time as a function of forgetness rate.}\label{transients}
\end{figure*}

After a certain time-of-flight the particles reach the loops and the interaction between $L$ and $AB$ defined in Eqs. \eqref{LAB1}, \eqref{LAB2} becomes important. 
At this point the relevant term in the wavefunction \eqref{decohered} proportional to $ \sqrt{p} $ evolves as
\begin{align}
\left(  \dfrac{\vert s \rangle_{AB}  \vert + \rangle_{E} +  \vert t \rangle_{AB} \vert - \rangle_{E}   }{\sqrt{2}}  \right) \vert \Phi_{0} \rangle_{L} \nonumber \mapsto \\ \nonumber
\\ 
 \vert s \rangle_{AB}  \left(  \dfrac{\vert + \rangle_{E} \vert \Phi_{0} \rangle_{L} +  \vert - \rangle_{E} \vert \Phi_{1} \rangle_{L} }{\sqrt{2}}  \right) 
\label{correction}
\end{align}
The particles $ A $ and $ B $ are corrected back to the singlet state, their mutual potential becomes once again attractive and they move towards each other and out of the loops. After a transient time, the orbital wave functions settle back in the equilibrium position of $ V(R) $ for the molecular term $ ^{1}\Sigma^{+} $ and the particles fall back in their initial singlet state. 
After tracing out the position degrees of freedom we can give an effective description of this dynamics in terms of a \textit{recovery} superoperator acting on the joint system $ ABL $. This recovery channel will be denoted $ \mathcal{R} $. 
Note that interaction of the particles with the environment leads to an increase in the entropy of both systems and to a loss of coherence in the particles' state, however, the unitary interaction between the particles and the loop exchanges the entanglement among the particles and the environment to entanglement among the loop and the environment, recovering the purity and phase to the particles' original singlet state.

After one recovery cycle the loop is entangled with the environment and has non-vanishing entropy. An external source of energy is required to erase the loop's state and reset it to $ \vert \Phi_{0} \rangle$, so that the whole process can start over. This erasing step requires a source of non-equilibrium and can happen stochastically. The non-equilibrium source could be a heat bath at a higher temperature, a strong magnetic field or any energetic resource floating around the environment. It is analogous to the sun in the case of photosynthesis or ATP for cell processes. Effectively we may describe this step by an appropriate amplitude damping channel we will refer to as the \textit{forgetness} map $ \mathcal{F} $.

\section{Master equation}

The behaviour of the qubot qualitatively described above is complex, certainly hard to predict algebraically and perhaps numerically challenging: it requires modelling involved stochastic quantum equations.
It is expected however that general features of the system can be effectively captured by a Markovian master equation describing a continuous limit in which the recovery and forgetness jumps happen on a time scale much shorter than the time over which the qubot is observed. The system is then expected to behave as a continuous error correcting code like the one discussed in \cite{Zurek1998}.

Dephasing $ \mathcal{D} $ competes with the recovery $ \mathcal{R} $ and forgetness $ \mathcal{F} $ and the joint $ ABL $ density matrix $ \rho \in \mathcal{B}(\mathcal{H}_{AB} \otimes \mathcal{H}_{L}) $ effectively evolves according to the master equation,
\begin{eqnarray}
\dot{\rho} = -i [H, \rho] + \mathcal{D} (\rho) + \mathcal{R} (\rho) + \mathcal{F} (\rho)
\end{eqnarray}
where $ H = \mathbb{1} \otimes \frac{\Delta }{2} Z$ is the loop Hamiltonian and the general Lindblad form $ \mathcal{L}(\rho) $ is given by
\begin{eqnarray}
\mathcal{L}(\rho)  = \sum_{a} \left( L_{a} \rho L_{a}^{\dagger} - \dfrac{1}{2} L_{a}^{\dagger} L_{a} \rho  -  \rho \dfrac{1}{2} L_{a}^{\dagger} L_{a} \right)
\end{eqnarray}
with $ L_{a} $ the quantum jump operators. Free Hamiltonian evolution of the spins has been omitted for simplicity and we take $ \hbar = 1 $.



\begin{figure*}[t] 
    \centering
    \includegraphics[width = \textwidth]{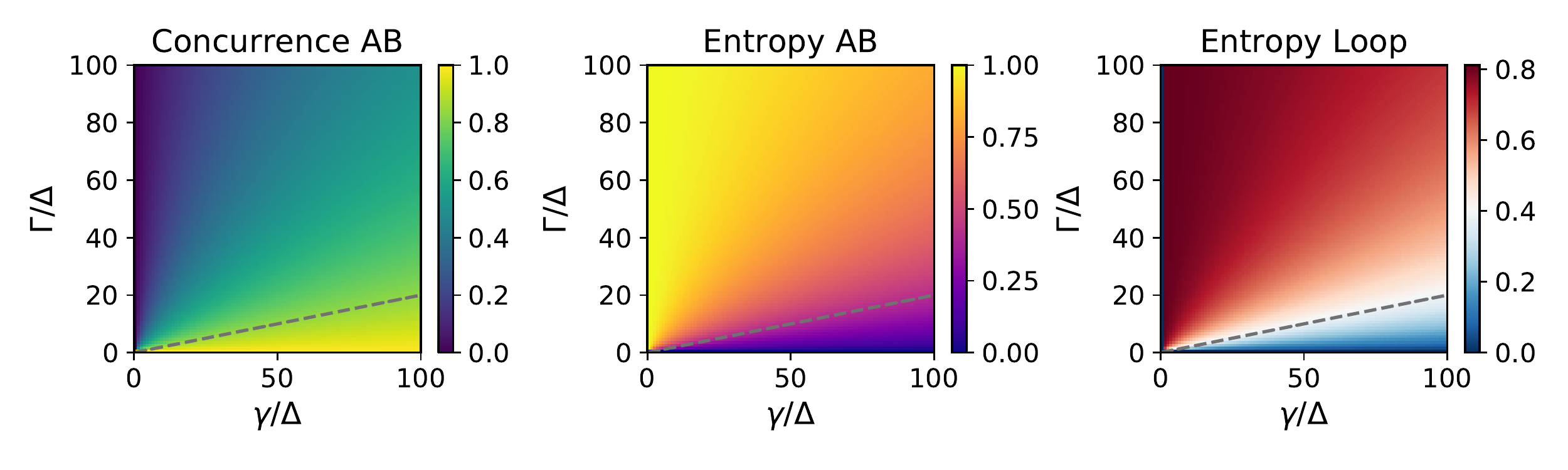}
    \caption[]{\small  Steady state concurrence of spins $ C(AB) $, entropy of spins $ S(AB) $ and entropy of loop $ S(L) $ as a function of the decoherence and forgetness rates $ \Gamma, \gamma $. Dashed lines indicate the curve $ \Gamma = 0.2 \gamma $, below which the system exhibits high entanglement and low entropies.
    }
    \label{color}
\end{figure*}

The system $ AB $ is comprised of two qubits, but we are interested in the two dimensional subspace generated by the singlet and triplet states $ \lbrace \vert s \rangle, \vert t \rangle \rbrace $. We will thus treat $ AB $ as a logical qubit. Recalling the definition of the computational basis $ \vert \uparrow \downarrow \rangle_{AB} \equiv \vert \bar{0} \rangle, \vert  \downarrow\uparrow \rangle_{AB} \equiv \vert \bar{1} \rangle  $ given above, the quantum jump operators for the dephasing channel are
\begin{eqnarray}
D_{0} = \sqrt{\Gamma} \vert \bar{0} \rangle \langle \bar{0} \vert \otimes  \mathbb{1}   \ , \ D_{1} = \sqrt{\Gamma}\vert \bar{1} \rangle \langle \bar{1} \vert \otimes  \mathbb{1} \ ,
\end{eqnarray}
where $ \Gamma $ is the decoherence rate.

The channel $ \mathcal{R} $ is defined by the recovery operators
\begin{eqnarray}
R_{0} = \sqrt{r} \vert s \rangle \langle s \vert \otimes  \mathbb{1}  \ , \ R_{1} = \sqrt{r} \vert s \rangle \langle t \vert \otimes X
\end{eqnarray}
where $ \vert s \rangle = \left(   \vert \bar{0} \rangle - \vert \bar{1} \rangle  \right) / \sqrt{2} $, $ \vert t \rangle = \left(   \vert \bar{0} \rangle + \vert \bar{1} \rangle  \right) / \sqrt{2} $, $ X = \vert \Phi_{0} \rangle \langle \Phi_{1} \vert +  \vert \Phi_{1} \rangle \langle \Phi_{0} \vert  $ is the bit-flip operator in the $ \lbrace \vert \Phi_{0} \rangle, \vert \Phi_{1} \rangle  \rbrace $ basis and $ r $ is the recovery rate. For our qubot, the recovery rate is bound by the inverse time necessary to complete one full recovery and return to the initial state. We can then say that $ r $ is roughly the inverse `recovery time' given by
\begin{eqnarray}
r \approx \left( t_{c} + \gamma^{-1} \right)^{-1}
\end{eqnarray}
where $ t_{c} $ is a correction time (that is the time needed for the particles $ AB $ to reach the loop plus the time necessary for the $ ABL $ interaction to occurr) and $ \gamma^{-1} $ the inverse forgetness rate, or the time needed for the loop to reset to state $ \vert \Phi_{0} \rangle $.

Finally, the $ \mathcal{F} $ superoperator consists in an amplitude damping channel and is hence given by the jump operator
\begin{eqnarray}
F = \sqrt{\gamma} \  \mathbb{1} \otimes  \vert \Phi_{0} \rangle \langle  \Phi_{1} \vert \ . \  
\end{eqnarray}


We can numerically integrate the master equation for the initial state 
\begin{eqnarray}
\rho(0) = \vert s \rangle \langle s \vert \otimes \vert \Phi_{0} \rangle \langle \Phi_{0} \vert
\end{eqnarray}
and compute the concurrence $ C(AB) $ and entropy $ S(AB) $ of $ AB $, and the entropy of the loop $ S(L) $ as functions of normalized time $ \Delta t $ (that is, time measured in units of the transition frequency $ \Delta^{-1} $ between states of the loop) \cite{qutip}. For comparison, we also calculate the concurrence of a pair of free spins initially in a singlet state under the influence of a dephasing channel with rate $ \Gamma $. Since $ r^{-1} $ is given by $ \gamma^{-1} $ plus an additive constant $ t_{c} $ defined solely by internal properties of the system, it is interesting to study the behavior of the qubot in terms of $ \Gamma $ and $ \gamma $. Hence we assume $ t_{c} \ll \gamma^{-1} $, the rate of available resources is much shorter than the reaction rate of the system.


As we can see in Figure \ref{transients}(a) the concurrence of a pair of free spins in a dephasing environment rapidly degrades to zero (blue curve).  The qubot spins, on the other hand, decay to a mixed steady state with non-vanishing entanglement (orange curve). The dashed red and green curves show the qubot and loop von Neumann entropies, respectively. Initially zero, the entropies increase to a steady value.
The levels of entanglement and entropy to which the system stabilizes and the necessary time for reaching stability are dependent on the decoherence $ \Gamma $, recovery $ r $ and forgetness $ \gamma $ rates.

An interesting question is how long the device takes to reach the steady state. To quantify that, we define the \textit{stabilization time} $ t_{o} $ as the time at which the concurrence of $ AB $ satisties the condition
\begin{eqnarray}
\dfrac{C_{t_{o}} - C_{\infty}}{C_{\infty}} = 0.1 \% \ , \ 
\end{eqnarray}
where $ C_{t_{o}} $ is the concurrence at time $ t_{o} $ and $ C_{\infty} $ the concurrence at infinite time. Numerically, $ C_{\infty} $  is reached when variations in the concurrence become of the order of the numerical precision.
A plot of the stabilization time is shown in Figure \ref{transients}(b) as a function of the recovery rate $ \gamma / \Delta $ for four different decoherence rates. For smaller values of $ \Gamma $ (weaker decoherence), small increments in $ \gamma / \Delta $ make a large difference in the stabilization time $ \Delta t_{o} $. As the decoherence rate assumes larger values (stronger decoherence), increasing $ \gamma / \Delta $  becomes less effective.

\begin{figure*}[t] 
    \centering
    \includegraphics[width = 0.9\textwidth]{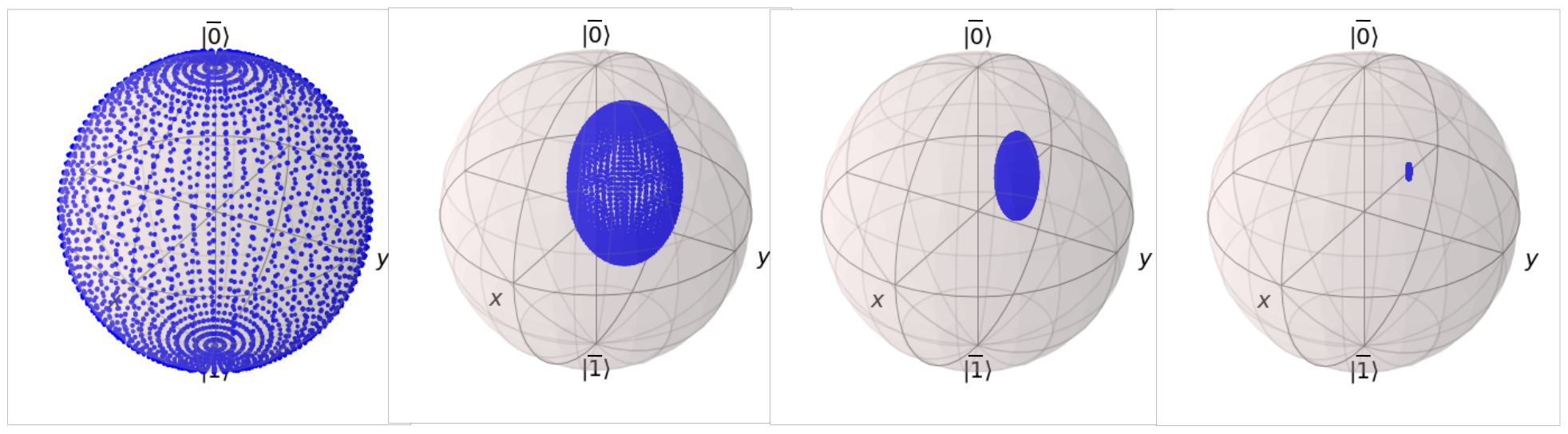}
    \caption[]{\small  Evolution of the logical Bloch sphere spanned by the states $ \vert \uparrow \downarrow \rangle_{AB} \equiv \vert \bar{0} \rangle, \vert  \downarrow\uparrow \rangle_{AB} \equiv \vert \bar{1} \rangle $: any initial state of the particles $ AB $ for which the spins are anti-parallel evolve to a mixed state close to the singlet. From left to right, $ \Delta t = 0.0, 0.4, 0.8, 2.0 $, respectively. Rates for this plot are $ \Gamma / \Delta = 1, \gamma/\Delta = r / \Delta = 1.5 $.
    }
    \label{bloch}
\end{figure*}

Figure \ref{color} shows the steady state values of concurrence and entropies for the qubot and loop as a function of the rates $ \Gamma, \gamma $. The dashed line indicates the curve $ \Gamma = 0.2 \gamma $, clear from the bare plots of concurrence and entropy. Below this line $ C(AB) \gtrsim 0.6 $ and $ S(AB), S(L) \lesssim 0.2 $, indicating a considerable amount of entanglement and a low entropy. We thus numerically find that if the decoherence and recovery rates satisfy,
\begin{eqnarray}
\gamma > 5  \Gamma
\label{operation_point}
\end{eqnarray}
the qubot is able to protect the e-bit at a significant value of concurrence. 
Erasing the loop state back to $ \vert \Phi_{0} \rangle $ is an operation which takes at least a time $ t_{erase} \approx \Delta^{-1} $ \cite{Lloyd2000}. Hence, the maximum forgetness rate is expected to be $ \gamma_{max} \approx \Delta $. In combination with \eqref{operation_point} we find that a successful qubot satisfies
\begin{eqnarray}
 \Delta \gtrsim 5 \Gamma 
 \label{operation_point2}
\end{eqnarray}
The lifetime for laboratory ultracold dipolar molecules  \cite{Rvachov2017} has been reported to be $ \tau \sim 200 \ \mu\mathrm{s}  = (5 \ \mathrm{kHz})^{-1} $. Among the decoherence processes that can destroy the molecule is the dephasing channel inducing singlet-triplet transitions, so the lifetime $ \tau_{D} $ of the molecule due to dephasing alone satisfies $ \Gamma^{-1} = \tau_{D} \geq \tau $, which implies $ \Gamma \leq \tau^{-1} $. We may then substitute condition \eqref{operation_point2} above by $ \Delta > 5 \tau^{-1} \sim 25 \ \mathrm{kHz} $.   
Assuming a typical superconducting quantum circuit energy on the scale of $ \Delta \sim 1 \ \mathrm{GHz} $, we see that the condition for operation of the qubot can in principle be satisfied within current experimental technology. The qubot can then be used to extend the molecule's lifetime to periods longer than the one set by the decoherence rate $ \Gamma $, and consequently protect the particles' spin singlet state.


What happens if the initial state of the particles $ AB $ is different from a singlet? Figure \ref{bloch} shows the time evolution of points on the logical Bloch sphere spanned by the states $ \vert \uparrow \downarrow \rangle_{AB} \equiv \vert \bar{0} \rangle, \vert  \downarrow\uparrow \rangle_{AB} \equiv \vert \bar{1} \rangle $. With the specification of interaction with the environment given in equation \eqref{decohered} it makes sense to consider states within this subspace. The north and south poles of the sphere correspond to the logical states  $\vert \uparrow \downarrow \rangle_{AB}$ and $\vert  \downarrow\uparrow \rangle_{AB}$ while the $ +\hat{x} $ and $ -\hat{x} $ directions correspond to the triplet and singlet states, respectively. It is possible to see that the current qubot example not only protects the singlet, but evolves any initial state on the logical Bloch sphere to a mixed state close to a singlet. 
To initialize the qubot, it is sufficient to prepare the particles' spins in any anti-parallel configuration.
For the parameters $ \Gamma / \Delta = 1, \gamma/\Delta = r / \Delta = 1.5 $ the steady state fidelity with respect to the singlet is $ F \approx 0.82 $. 

It is important to note that the current qubot example stabilizes a preferred state, namely the singlet. This is analogous to the so-called quiescent state in the surface code, for which only a preferred state is protected and there are no free degrees of freedom if the stabilizer operators are enforced on all physical qubits \cite{Fowler2012}. As in the surface code, however, nothing prevents one from designing more general qubots stabilizing complex quantum states. An interesting open question is how to engineer a qubot capable of protecting an ensemble of logical qubits in an arbitrary quantum state. 

\section{Discussion}

A common feature of living organisms is employing different molecular conformations to manipulate a potential barrier and consequently execute a function.
Na, K ATPase ion pumps provide an interesting example in which conformations alter the potential landscape seen by an ion and consequently transport it in a preferred direction \cite{Astumian1998, Gadsby2009}.
The presented qubot example operates in a similar way exploiting interatomic potential changes and moving parts to ``proofread'' and protect quantum information, consequently sustaining a preferred quantum state, namely the singlet.  

In the future, nano-mechanical designs with moving parts could be built using single atom tweezers \cite{Bruno2019, Cooper2018} and the available ultracold molecule toolbox \cite{Rvachov2017, Park2017, Sikorsky2018, Ni2018, Ding2020}.
Trapped ions \cite{Kielpinski2012}, NV-centers \cite{Marcos2010} and levitated nano-particles \cite{Martinetz2020, Conangla2018} can also couple internal and motional degrees of freedom to superconducting circuits, and similar mechanisms can be envisioned.
One can also imagine qubots with no moving parts where quantum errors induce a potential change which triggers a recovery operation involving only internal states of the system. For instance, a spin-ensemble quantum memory of NV-center defects in diamond glued on top of a superconducting circuit can collectively couple to a qubit \cite{Kubo2011}. This could be exploited to create a self-error correcting system capable of protecting quantum data stored in the spins. 
\\

\paragraph{Acknowledgments \textemdash } The author acknowledges discussions with Lucianno Defaveri, Bruno Melo, and George Svetlichny. In 2019 the author attended the \textit{Prospects in Theoretical Physics} program at the Institute for Advanced Studies in Princeton. The meeting was centered on ``Great Problems in Biology for Physicists'' and had an important impact in the development of this work. This work was financed in part by Coordena\c{c}\~ao de Aperfei\c{c}oamento de Pessoal de N\'ivel Superior - Brasil (CAPES) - Finance Code 001 and by Conselho Nacional de Desenvolvimento Cient\'ifico e Tecnol\'ogico (CNPq) and by the FAPERJ Scholarship No. E-26/202.830/2019.

\onecolumngrid
\pagebreak
\begin{center}
\textbf{\large Appendix: an alternative environment}
\end{center}
\setcounter{equation}{0}
\setcounter{figure}{0}
\setcounter{table}{0}
\makeatletter
\renewcommand{\theequation}{S\arabic{equation}}
\renewcommand{\thetable}{S\arabic{table}}
\renewcommand{\thefigure}{S\arabic{figure}}

\begin{figure*}[h!] 
    \centering
    \begin{subfigure}[b]{0.35\textwidth}
    \centering
        \includegraphics[width=\textwidth]{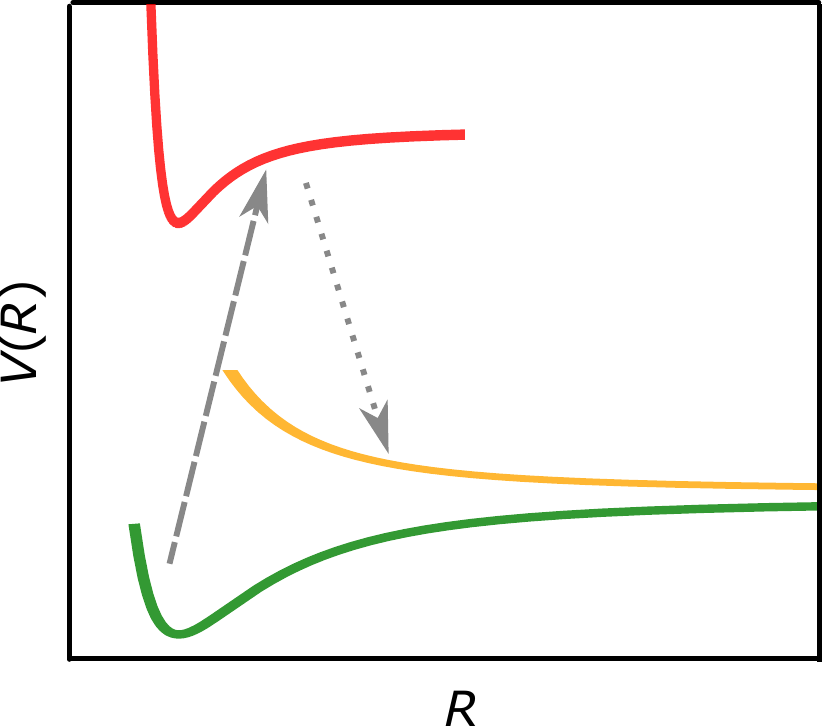}
        \caption{ \ }
        \label{fig:gull}
    \end{subfigure}
    ~ 
    \begin{subfigure}[b]{0.4\textwidth}
    \centering
        \includegraphics[width=\textwidth]{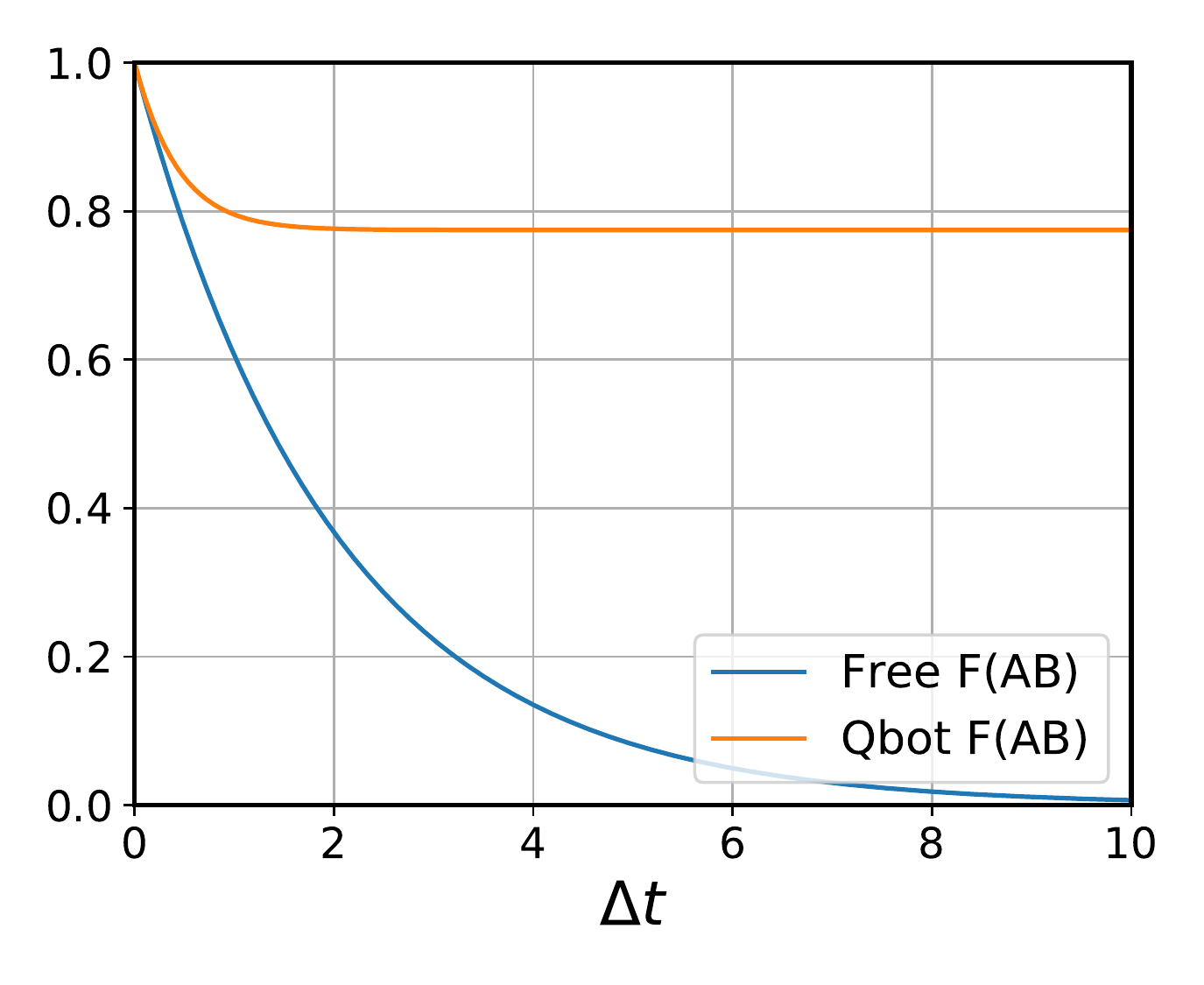}
        \caption{ \ }
        \label{fig:tiger}
    \end{subfigure}
    \caption[]{\small  (a) Schematic representation of the dissociation process. (b) Time-dependent fidelity with respect to singlet; rates for this plot are $ \Gamma / \Delta = 1, \gamma/\Delta = r / \Delta = 1.5 $.}\label{dissociation}
\end{figure*}

In the main text we have considered the dephasing environment which ``learns'' with a certain probability the spin states of particles $ AB $ thus causing decoherence of the e-bit. 
We might consider different environments. A case of interest for ultracold quantum chemistry experiments is ``dimer destruction by a diode laser'' as studied in \cite{Ban2005}. 
A molecule is formed by particles $ AB $, bound in the $ ^{1}\Sigma^{+} $ molecular term. 
A laser shining at the molecule may induce a transition to an excited state, and subsequent decay to either the $ ^{1}\Sigma^{+} $ or $ ^{3}\Sigma^{+} $ states. When the system decays to the $ ^{3}\Sigma^{+} $ state, the potential between $ AB $ changes to repulsive, and the molecule is destroyed. This is called photodissociation and a schematic representation of the process is shown in Figure \ref{dissociation}(a). In this situation the spins are still in a triplet entangled state after the molecule has dissociated, however, the particles will fly apart and the e-bit will be delocalized and possibly lost in practice. 

The ``photodissociation environment'' $ \mathcal{P} $ given by the laser acts as  
\begin{align}
\vert s \rangle_{AB} \vert 0 \rangle_{E} \vert \Phi_{0} \rangle_{L} &\,\mapsto \sqrt{1 - p} \vert s \rangle_{AB} \vert 0 \rangle_{E} \vert \Phi_{0} \rangle_{L}  + \sqrt{p} \vert t \rangle_{AB} \vert 1 \rangle_{E} \vert \Phi_{0} \rangle_{L}
\end{align}
where $ p$ is the probability of error and $ \vert 0 \rangle_{E} , \vert 1 \rangle_{E}$ are orthogonal states of the environment. The qubot corrects this state to
\begin{align}
    \vert s \rangle_{AB} \otimes (\sqrt{1 - p} \vert 0 \rangle_{E} \vert \Phi_{0} \rangle_{L}  + \sqrt{p} \vert 1 \rangle_{E} \vert \Phi_{1} \rangle_{L})
\end{align}
thus entangling the loop to the environment. The forgetness map then erases the loop memory and the process restarts.

If the molecule transitions to the triplet state, dissociation prevents the system from going back to the singlet configuration. We can describe this process as a decay from $ \vert s \rangle $ to $ \vert t \rangle $. The corresponding quantum jump operator is $ P = \sqrt{\Gamma} \vert t \rangle \langle s \vert \otimes \mathbb{1} $.
We numerically solve the master equation presented in the main text, with the superoperator $ \mathcal{P} $ replacing $ \mathcal{D} $. Figure \ref{dissociation}(b) shows the fidelity between the original singled spin state and the time-dependent state under the action of photodissociation for free spins (blue curve) and the qubot (orange curve).

\end{document}